\documentclass[twocolumn,showpacs,prb]{revtex4}

\usepackage{graphicx}
\usepackage{dcolumn}
\usepackage{bm}

\begin{document}

\title{Diffusion-reaction mechanisms of nitriding species in SiO$_2$}
\author{W. Orellana} 
\email{wmomunoz@if.usp.br} 
\author{Ant\^onio J. R. da Silva}
\author{A. Fazzio}
\affiliation{Instituto de F\'{\i}sica, Universidade de S\~ao Paulo, 
C.P. 66318, 05315-970, S\~ao Paulo, SP, Brazil}

\date\today

\begin{abstract} 
We study using first-principles total-energy calculations, diffusion-reaction
processes involved in the thermal nitridation of SiO$_2$. We consider NO, NH, 
N$_2$ and atomic N in different charge states as the nitriding species 
in $\alpha$-quartz. 
Our results show that none of neutral species react with the SiO$_2$ network
remaining at interstitial sites. Therefore, they are likely to diffuse through 
the oxide, incorporating nitrogen at near-interface (Si/SiO$_2$) regions. 
Whereas, charged species are trapped by the network, nitriding bulk SiO$_2$.
For the diffusing species, 
we find that NH and atomic N show increasing diffusivities with temperatures, 
whereas for NO and N$_2$ they are relatively constant. This result agree well 
with the finding of higher N concentration at the Si/SiO$_2$ interface obtained
by thermal processing of SiO$_2$ in NH$_3$ as compared with those obtained
in N$_2$O. Finally, we discuss spin-dependent incorporation reaction mechanisms 
of NH and atomic N with the SiO$_2$ network.

\end{abstract} 

\maketitle

\section{Introduction}

Nitrided silicon oxide or oxynitride is currently the near-term solution 
to substitute SiO$_2$ as the gate insulator material for the ultrathin 
metal-oxide-semiconductors (MOS) technology.
Recent studies have suggested that the performance of oxynitride-based MOS 
devices depends both on the depth concentration and on the distribution of 
nitrogen into the gate oxide.\cite{green}
According to these studies, the best nitrogen profile for an ultrathin gate
oxide would have: 
(i) a small nitrogen concentration near the SiO$_2$/Si interface in order to 
reduce degradations by hot electrons,\cite{cartier} and 
(ii) a larger N concentration near the interface between the dielectric and
the polycrystalline silicon (poly-Si) gate electrode, in order to minimize 
dopant diffusion.\cite{wrister}

The growth of ultrathin oxynitride films depends strongly on the reactant 
agent ({\it e.g.}, N$_2$O, NO, NH$_3$) and the technique used. 
Nitrogen can be incorporated into SiO$_2$ using either thermal oxidation
and annealing or chemical and physical deposition methods. 
Thermal nitridation of SiO$_2$ in N$_2$O generally results in a relatively 
low N concentration at near-interface (Si/SiO$_2$) region. The nitriding
species are originated in the decomposition of the N$_2$O molecule occurring 
at typical oxidation temperature, these species being the NO and N$_2$ 
molecules.\cite{ellis,baumvol,lu} 
On the other hand, nitrogen incorporation in SiO$_2$ can also be performed
via annealing in a NH$_3$ atmosphere, resulting in a relatively high N 
concentrations into the films.\cite{baumvol}
This method provides both near-interface and near-surface nitridation, which
suggests different mechanisms for N incorporation or different nitriding 
species derived from NH$_3$. The above thermal processing are performed 
at high temperatures ($>$\,800~$^{\circ}$C).
Plasma nitridation is a promising method for making ultrathin oxynitride
at lower temperatures ($\sim$\,300--400~$^{\circ}$C).
Higher N concentrations and controlled distribution can be attained with
this method,\cite{watanabe} typically using ions and radicals derived from 
N$_2$ and NH$_3$ as nitrogen sources.\cite{gusev} Additionally,
nitridation by energetic particles (N ions) provides high 
N concentrations much closer to the near-surface region,
with little or no nitrogen at the SiO$_2$/Si interface.\cite{green}
Although the nitridation mechanisms to make ultrathin SiO$_2$ films are well
known at the few layer level, less is known about the diffusing mechanism of 
the nitriding species and their reactions with the oxide at the atomic level.

In this work the energetics and diffusing properties of the SiO$_2$ thermal
nitridation are studied from first-principles total-energy calculations. 
We have considered the N$_2$, NH and NO molecules as well as atomic N in 
different charge states as the nitriding species reacting and diffusing 
through the SiO$_2$ network. 
The outline of the paper is as follows. Section II describes the theoretical
procedure. Our results for the interaction of the nitriding species with the
SiO$_2$ network is presented in Sec.~IIIA. The diffusivity of the nitriding 
species through SiO$_2$ is presented in Sec.~IIIB. In Sec.~IIIC we discuss the
spin-dependent diffusion reactions of NH and atomic N with the SiO$_2$ network.
Finally, in Sec.~IV we show our conclusions.

\section{theoretical method}

Our calculations were performed in the framework of the density functional 
theory,\cite{dft} using a basis set of numerical atomic orbitals as 
implemented in the SIESTA code.\cite{siesta}
We have used a split-valence double-$\zeta$ basis set plus the polarization 
functions as well as standard norm-conserving pseudopotentials.\cite{pseudo} 
For the exchange-correlation potential we adopt the generalized gradient 
approximation.\cite{pbe}
We used a 72-atom $\alpha$-quartz supercell and the $\Gamma$ point for the
Brillouin zone sampling. 
The positions of all the atoms in the supercell were relaxed until all the 
force components were smaller than 0.05 eV/$\text{\AA}$.
We also consider neutral and singly charged species, where the neutrality 
of the cell is always maintained by introducing a compensating background 
charge. Spin-polarization effects are included throughout the calculation
since they are important for the correct description of atomic and molecular 
reaction processes in SiO$_2$.\cite{prl2-us}
We initially study the energetics and structural properties of N$_2$, NH, NO, 
and atomic N in the largest interstitial site of $\alpha$-quartz 
for three different charge states ($+, 0, -$). We explore possible 
reactions that these species may undergo with the network as well as 
their diffusivities in $\alpha$-quartz.

\section{results and discussion}

\subsection{Incorporation of nitriding species into the SiO$_2$ network}
 
\begin{figure} 
\includegraphics[width=7cm]{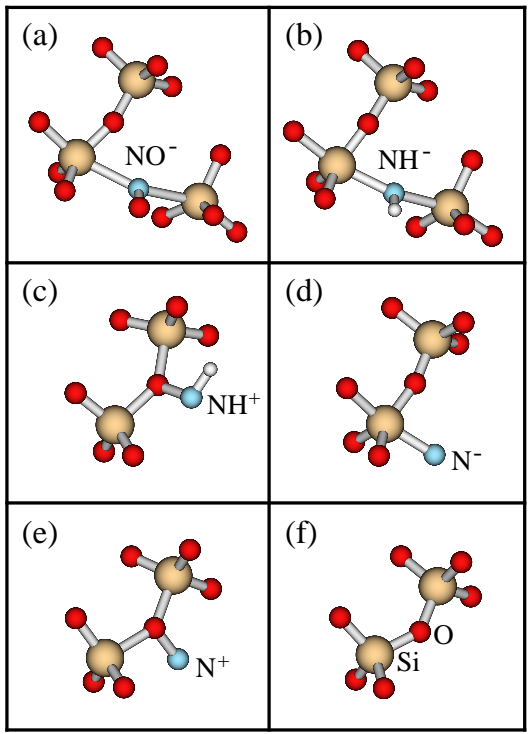}
\caption{\label{f1} Local equilibrium geometries for singly 
charged nitriding species after reacting with the SiO$_2$ network.
(a) negative NO.
(b) negative and (c) positive NH, respectively.
(d) negative and (e) positive atomic N, respectively.
(f) the perfect SiO$_2$ network.}
\end{figure}

Thermal processing of SiO$_2$ in N$_2$O and NO atmospheres shows that the 
NO molecule is the species which diffuses through the oxide, being 
incorporated at the Si/SiO$_2$ interface.\cite{lu}
In our calculations we find that neutral NO and NO$^{+}$ do not react 
with the SiO$_2$ network, remaining at the interstitial position and being, 
in this way, able to diffuse through the oxide as experimentally observed.
However, NO$^{-}$ is trapped by the network forming a structure 
where the N atom of the molecule is bound to two fourfold coordinate 
Si atoms of SiO$_2$, as shown in Fig.~\ref{f1}(a). 
The binding energy of NO$^{-}$, calculated as the difference in energy 
between the bound and the free-interstitial configurations for the same
charge state, is found to be 1.3~eV. 
The Si-N and N-O bond lengths are 2.26 and 1.23~\AA, respectively, 
whereas the Si-Si distance decreases by about 6\% with respect to the 
equilibrium SiO$_2$ structure. 
For interstitial NO and NO$^{+}$ the N-O bond lengths are 1.17 and 1.13~\AA, 
respectively.

The NH radical seems to be the most likely diffusing species after thermal 
dissociation of NH$_3$ according to the dissociative reactions:
NH$_3$ $\rightarrow$ NH$_2$\,+\,H\,$-$\,4.5~eV and
NH$_2$ $\rightarrow$ NH\,+\,H\,$-$\,3.9~eV.
Although NH$_2$ may be another diffusing species for the interface nitridation,
it is a relatively larger molecule and it may suffer additional dissociations
at near-surface SiO$_2$ during thermal processing. According to our results,
neutral NH in the gas phase has a spin triplet ($S$\,=1) ground state whereas 
the singlet ($S$\,=0) states is 2.06~eV higher in energy. 
However, inside the largest interstitial site of $\alpha$-quartz, this 
difference in energy between both spin configurations decreases to 1.04~eV, 
due to the interaction with the crystal field.
We find that triplet NH does not react with the SiO$_2$ network remaining 
relatively inert at the interstitial sites. However, singlet NH reacts quickly 
with the oxide forming a structure where the N atom of NH binds to both
Si and O atoms of the oxide. This incorporation reaction is highly 
exothermic with an energy gain of 2.04~eV. 
In Section III C we discuss the possibility that NH might be incorporated 
in the SiO$_2$ network via a triplet-to-singlet spin exchange.

Singly charged NH molecules are highly reactive in the SiO$_2$ network 
forming bound configurations as shown in Fig.~\ref{f1}(b) and Fig.~\ref{f1}(c). 
The binding energies for NH$^{-}$ and NH$^{+}$ are 
3.2 and 2.7~eV, respectively, which suggest that these configurations 
are very stable favoring a SiO$_2$ near-surface nitridation. We observe 
that the NH$^{-}$ bound configuration is similar to that previously found 
for NO$^{-}$. 
However, the relaxation of the SiO$_2$ network is larger for the NH$^{-}$
bound configuration [see Fig.~\ref{f1}(b)]. Here, both Si-N bond lengths are 
about 2.1~\AA, whereas the Si-Si distance decreases by about 14\% with respect 
to the perfect SiO$_2$ showing a strong lattice relaxation. The N-H bond 
length is found to be 1.04~\AA. 
In the bound NH$^{+}$ configuration [Fig.~\ref{f1}(c)], the N atom binds to
an O atom of the oxide keeping its bond to the H atom. The O-N and 
N-H bond lengths are 1.43 and 1.05~\AA, respectively, forming a O-N-H angle 
of 103$^{\circ}$.

The N$_2$ molecule is a product of the N$_2$O gas decomposition at typical
oxidation temperatures, being introduced in this way into SiO$_2$ in thermal 
processing.\cite{green} N$_2$ (and atomic N) may also be introduced at lower 
temperatures by plasma assisted methods.\cite{watanabe}
Our results for N$_2$ in SiO$_2$ show that this molecule does not react with 
the oxide being relatively inert for the three charge states considered. 
Therefore, it may diffuse easily through the oxide reacting with the silicon 
at the Si/SiO$_2$ interface or escaping from the films.

The N atom in free space has a quartet ($S$\,=3/2) ground state.
The difference in energy with respect to the doublet 
($S$\,=1/2) state is calculated to be $\sim$\,3~eV. However, for a N atom 
in the largest interstitial site of $\alpha$-quartz, this energy difference
decreases to 0.76~eV.
We find that quartet N does not react with the SiO$_2$ network suggesting 
that it would be a diffusing species.
However, doublet N reacts with the oxide being incorporated into SiO$_2$ 
network forming a Si-N-O bond. This reaction is exothermic with an energy 
gain of 0.76~eV. In Section III C, we discuss the possibility that atomic 
N might be incorporated in the SiO$_2$ network via a quartet-to-doublet
spin exchange. 

Singly charged N atoms are highly reactive in SiO$_2$. N$^{-}$ binds
to a fourfold coordinate Si atom forming an additional Si-N bond of 
1.85~\AA, as shown in Fig.~\ref{f1}(d). The N atom has a binding energy 
of 2.2~eV. On the other hand, N$^{+}$ binds to an oxygen forming a O-N 
bond, as shown in Fig.~\ref{f1}(e). We find a O-N bond length of 1.39~\AA\ 
with a N binding energy of 1.1~eV.

As a general trend, we can say that because of the ionic character of SiO$_2$, 
charged species are trapped by the network. The positively (negatively) 
charged NO, NH and N species attach to O (Si) atoms of the oxide, 
forming strong bonds, with binding energies ranging from 1 to 3~eV. This 
suggests a mechanism for the high-density bulk and near-surface nitridation.
On the other hand, the neutral species in their ground-state spin 
configurations do not react with the SiO$_2$ network. Therefore, they are 
able to diffuse through the oxide.

\subsection{Diffusion of nitriding species in SiO$_2$}
 
\begin{figure} 
\includegraphics[width=8.5cm]{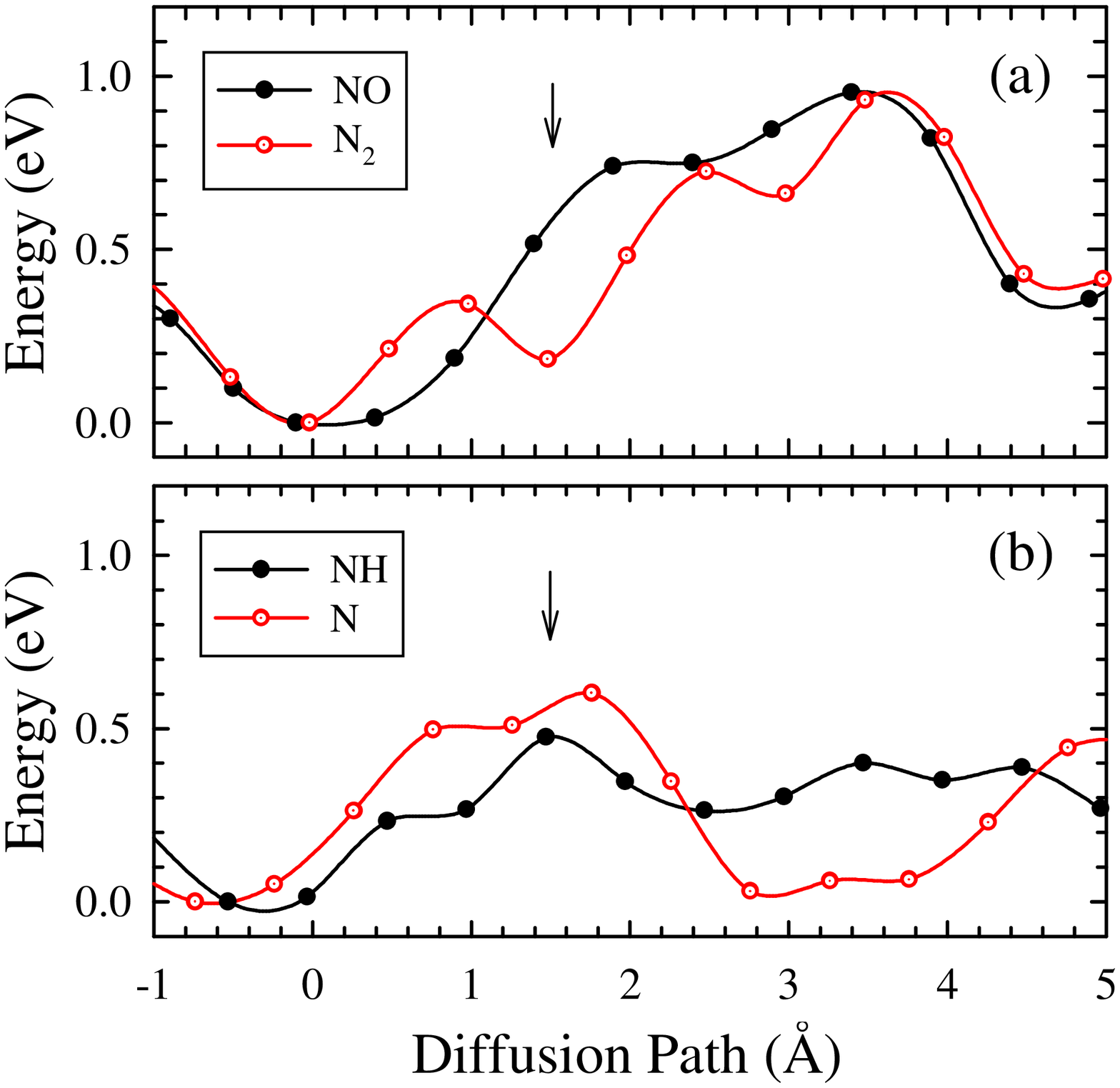}
\caption{\label{f2} Total energy variations for the neutral nitriding 
species diffusing through a channel normal to the $c$-axis of
$\alpha$-quartz.
(a) for the NO and N$_2$ molecules.
(b) for the NH molecule and atomic N.
The arrows indicate the point where the nitriding species pass closer
to the SiO$_2$ network and the zero distance is the center of the 
largest void in $\alpha$-quartz.}
\end{figure}

We calculate the diffusion coefficient for the neutral species hopping
through the larger channel of $\alpha$-quartz, normal to the $c$-axis.
To calculate the migration barriers, we have followed the same procedure 
as described in our previous calculation of the O$_2$ diffusion in 
$\alpha$-quartz.\cite{prl1-us} We fix one atom of the molecules at several
points along the pathway joining neighboring interstitial sites of 
$\alpha$-quartz where the distance between them is $\sim$\,5~\AA. 
All other atoms of the systems are allowed to relax. Our results 
for the total energy variations along this diffusion path are shown 
in Fig.~\ref{f2}. 
Here, the zero distance indicates approximately the center of the largest 
interstitial site and the arrow, the place of closest approach to the 
network.  As a general trend, we find an anisotropic energy profile through 
the migration path which can be associated to the asymmetry of neighboring 
interstitial sites.
Fig.~\ref{f2}(a) shows the calculated total-energy variations of diffusing 
NO and N$_2$; we observe that these species show similar global minima and 
height barriers along the pathway. However, N$_2$ exhibits 
two local minima not found in the NO pathway, showing that the interactions
between the crystal field and the molecular orbitals of the diffusing species 
may affect the diffusion energy profile. Nevertheless, we must consider that
uncertainties are introduced by fixing one atom of the molecules in order to 
map the diffusion energy profiles.
For NH and atomic N we find lower-energy barriers and shorter distances 
between minima, as compared with NO and N$_2$ (see  Fig.~\ref{f2}), 
suggesting higher diffusivities for these species.

We estimate the thermal diffusion coefficient or diffusivity for the nitriding
species using the Arrhenius law, given by 
\begin{equation}
D(T)=D_0\,\exp(-E_a/kT\,). 
\label{e1}
\end{equation}
Here, the prefactor is defined by $D_0$\,=\,$l{^2}\nu$/6 for a 
three-dimensional migration path, where $l$ is the 
hopping distance between minima and $\nu$ is the attempt frequency, $kT$ 
the Boltzmann constant times the temperature, and $E_a$ is the highest
energy barrier of the migration path. $\nu$ is calculated from the energy 
profile of each species at the interstitial site (approximately the zero 
distance in Fig.~\ref{f2}), using the harmonic approximation, whereas $E_a$ 
and $l$ are obtained directly from the same figure. 
\begin{table}[b]
\caption{\label{tab} Energy barriers ($E_a$), hopping distance between minima
($l$) and attempt frequencies ($\nu$) as obtained from the migration energy
profiles of Fig.~\ref{f2}. $D_0$ is the calculated pre\-factor coefficient of 
the diffusivities for each nitriding species.}
\begin{ruledtabular}
\begin{tabular}{ccccc}
Species & $E_{a}$(eV) & $l$(\AA) & $\nu$($\times$10$^{12}$sec$^{-1}$) & 
$D_{0}$($\times$10$^{-4}$cm$^{2}$/sec)\\
\hline \\
NO      & 0.95 & 4.7 & 2.26 & 8.32  \\
N$_{2}$ & 0.95 & 4.8 & 2.38 & 9.14  \\
NH      & 0.48 & 2.8 & 3.67 & 4.79  \\
N       & 0.60 & 3.6 & 3.41 & 7.37  \\
\end{tabular}
\end{ruledtabular}
\end{table}
Tab.~\ref{tab} lists the values of these quantities as well as our 
results for the prefactor of each nitriding species.
It is worth pointing out that we have considered a very specific migration
pathway through $\alpha$-quartz. Other pathways may result in different 
migration barriers and diffusivities. In fact, recent calculations have shown
major differences in the diffusivity of atomic hydrogen in different 
crystalline structures of SiO$_2$.\cite{bongiorno,tuttle} However, we 
find that our approach is valid to obtain informations on the relative 
diffusivities of the species under consideration. 

\begin{figure}
\includegraphics[width=8.5cm]{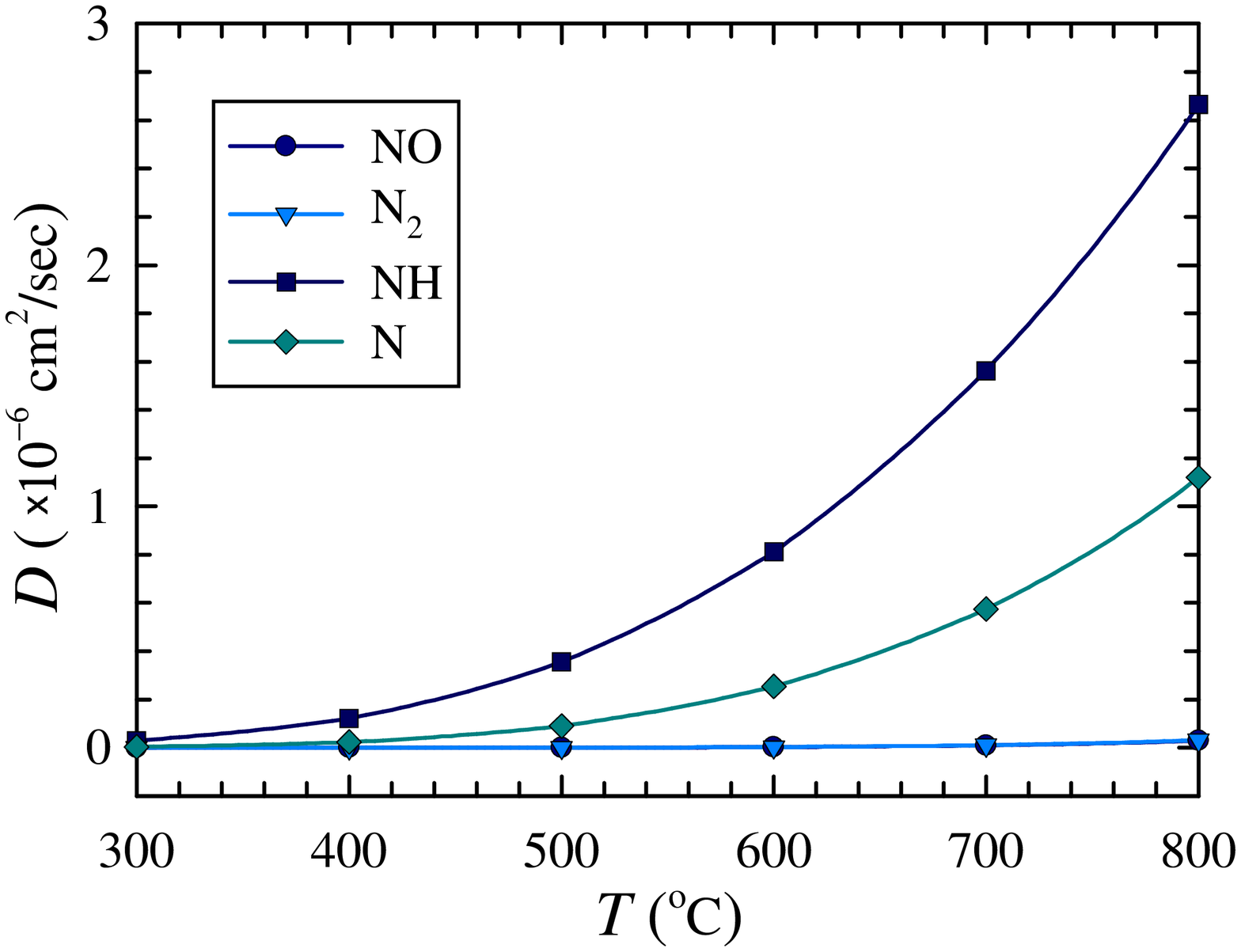}
\caption{\label{f3} Calculated diffusivities ($D$) of the nitriding 
species in $\alpha$-quartz as a function of typical plasma assisted and 
thermal processing temperatures.}
\end{figure}

Figure~\ref{f3} shows the diffusivities of the nitriding species in 
$\alpha$-quartz as calculated from Eq.~\ref{e1}, for a range of temperatures 
typically used in thermal and plasma assisted methods.
Here we observe that NH and atomic N have the higher diffusivities in 
$\alpha$-quartz with increasing temperatures as compared with NO and N$_2$, 
which suggest that they would be the best thermally-activated diffusing 
species in SiO$_2$ and the most efficient for Si/SiO$_2$ interface 
nitridation, as compared with NO and N$_2$.
According to our results, the ratio between NH and NO diffusivities at 
800$^\circ$C is estimated to be $D_{\rm NH}/D_{\rm NO}\simeq90$.
Assuming that NH and NO are the main diffusing species, our results agree 
well with the experimental finding of higher N concentrations at the 
interface obtained by thermal nitridation of SiO$_2$ in NH$_3$ as compared
with thermal nitridation in NO.\cite{baumvol}

\subsection{Spin-dependent diffusion reactions of NH and N in SiO$_2$}

\begin{figure}
\includegraphics[width=8.5cm]{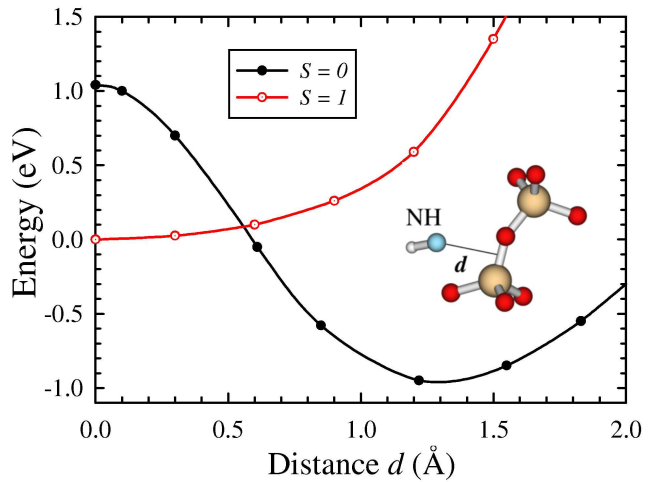}
\caption{\label{f4} Triplet ($S=1$) and singlet ($S=0$) 
total-energy curves along the reaction pathway ($d$ in the inset 
figure) for the NH molecule approaching the SiO$_2$ network.
The zero distance is the equilibrium position of NH in the largest
interstitial site of $\alpha$-quartz where its corresponding total
energy is chosen as the zero energy of the system.}
\end{figure}

As mentioned above, NH in the triplet state does not react with the SiO$_2$ 
network remaining inert at interstitial sites. However, singlet NH reacts 
easily with the oxide being incorporated into the network.
Because of this spin-dependent reaction of NH with the SiO$_2$ network, 
we study the possibility that a diffusing NH may suffer a triplet-to-singlet 
spin conversion while approaching the oxide, resulting in its incorporation 
into the network. 
For this, we compute the potential-energy surface (PES) along a pathway 
joining the equilibrium position of interstitial NH and a Si-O bond of 
the SiO$_2$ network, for both triplet and singlet spin states, as 
shown in Fig.\ref{f4}. The triplet PES in the figure is depicted along a
direction perpendicular to the diffusion pathway shown in Fig.~\ref{f2},
however, both pathways intercept at the interstitial equilibrium configuration
of NH (zero distance in Fig.~\ref{f4}).
The probability that the system changes from the triplet to the singlet 
PES may be estimated by the Landau-Zener theory.\cite{zener} 
This theory has been recently used to describe spin effects in the adsorption 
of O$_2$ in the Si(001) surface\cite{kato} and the O$_2$ incorporation 
reaction in the Si/SiO$_2$ interface.\cite{prl2-us}
Considering that the NH is initially in the triplet diffusion PES, and is 
evolving towards the crossing region with a velocity $v$, the probability 
for a conversion to the singlet PES ($P_{ts}$) may be approximated by
\begin{equation}
P_{ts}=2\,[\,1-\exp(-V^2/hv|F_1-F_2|\,)\,],
\label{e2}
\end{equation}
where $V$ is the triplet-to-singlet spin-orbit matrix element of NH 
($X\,^{3}\Sigma^{-}\!\!\rightarrow b^{1}\Sigma^{+}$) of
65.1~cm$^{-1}$,\cite{yarkony} $F_1$ and $F_2$ are the slopes of 
the two PES at the crossing point and $h$ is the Planck's constant. 
$v$ is estimated from the NH center-of-mass thermal velocity at 
800 $^{\circ}$C, a typical annealing temperature. $F_1$ and $F_2$ 
are obtained from the triplet and singlet curves at the crossing 
region (see Fig.~\ref{f2}). 
Thus, the probability for a triplet-to-singlet conversion is found to be
$P_{ts}$ = 9.3$\times$10$^{-4}$. This means that triplet NH in a single 
passage through the crossing point has a small probability to change to 
the singlet PES. However, as the system is thermally activated, many passages
of NH  through the crossing point will be performed.
In order to quantify the spin-conversion events for NH molecules approaching 
the SiO$_2$ network at a given temperature, we estimate the ratio between
the rate of NH following the triplet diffusion PES ($\Gamma_t$) and the rate
of NH changing from triplet to singlet PES at the crossing point 
($\Gamma_{ts}$). Assuming that the thermally-activated systems follow the 
Arrhenius law, the reaction rates can be written as
\begin{equation}
\Gamma_t=\Gamma_0\,\exp(-\Delta E_t/kT)
\label{e3}
\end{equation}
and
\begin{equation}
\Gamma_{ts}=\Gamma_0\,[\,\exp(-\Delta E_p/kT)\,]\!\times\!P_{ts}.
\label{e4}
\end{equation}
Here, $\Gamma_0$ is the frequency prefactor; $\Delta E_t$ is the energy 
barrier for the triplet diffusion PES obtained from Fig.~\ref{f2}(b)
($\Delta E_t$\,=\,0.5~eV) and $\Delta E_p$ is the energy difference between 
the lowest-energy equilibrium configuration of the triplet NH and the 
crossing point, obtained from Fig.~\ref{f4} ($\Delta E_p$\,=\,0.1~eV). 
For a temperature of 800 $^\circ$C, we obtain for our calculated $P_{ts}$ 
that $\Gamma_t/\Gamma_{ts}\simeq14$, which indicates that NH will proceed 
preferentially along the triplet diffusion PES.

As the incorporation of NH into the oxide due to the spin exchange is 
also likely to occur (an average of one in 14 events), we study three 
possible thermally-activated diffusion-reaction mechanisms for an 
incorporated singlet NH at 800 $^\circ$C:
(i) {\it NH diffuses along the SiO$_2$ network by a hopping mechanism}. 
The rate is estimated by Eq.~\ref{e3}, substituting $\Delta E_t$ by a 
calculated hopping energy barrier ($\Delta E_h$\,=\,2.2~eV). 
We find a hopping rate of $\Gamma_h$\,=\,$\Gamma_0$(4.6$\times$10$^{-11}$). 
(ii) {\it NH is released from the network following the singlet PES}.
This reaction rate is estimated by 
\begin{equation}
\Gamma_{s}=\Gamma_0\,[\,\exp(-\Delta E_{s}/kT)\,]\!\times\!(1-P_{ts}),
\label{e5}
\end{equation}
where $\Delta E_s$ is the singlet PES energy barrier obtained from 
Fig.~\ref{f4} ($\Delta E_s$\,=\,2.0~eV). 
We find that $\Gamma_s$\,=\,$\Gamma_0$(4.0$\times$10$^{-10}$). 
(iii) {\it NH is released from the network changing from the singlet to the 
triplet PES at the crossing point}. The reaction rate is estimated by
Eq.~\ref{e4}, where $\Delta E_{p}$ is now the difference between the energies
at the singlet-triplet crossing point and at the incorporated NH equilibrium
configuration, obtained from Fig.~\ref{f4} ($\Delta E_{p}$\,=\,1.05~eV).
We find that $\Gamma_{st}$\,=\,$\Gamma_0$(1.1$\times$10$^{-8}$). 
According to these results, the most likely diffusion mechanism for an
incorporated singlet NH in SiO$_2$ would be mediated by the reaction (iii),
i.e., NH leaves the network changing from the singlet to the triplet PES at 
the crossing point and then diffuses through SiO$_2$ as a triplet NH. 
The two other mechanisms are less likely, as can be verified by the ratios 
$\Gamma_{st}/\Gamma_h\simeq240$ and $\Gamma_{st}/\Gamma_s\simeq28$.

\begin{figure}
\includegraphics[width=8.5cm]{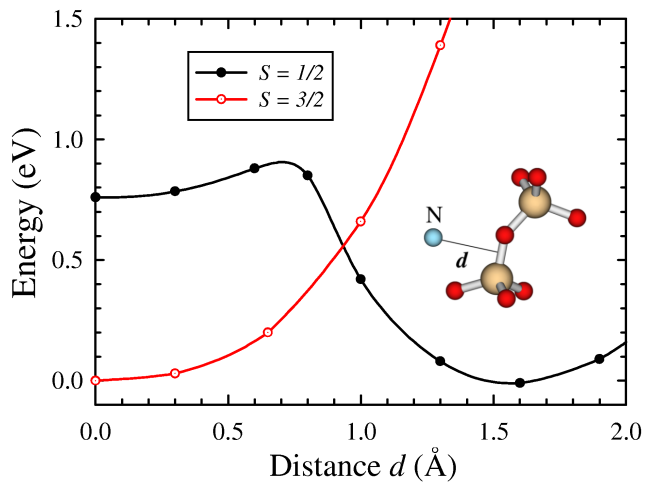}
\caption{\label{f5} Doublet ($S=1/2$) and quartet ($S=3/2$) total-energy 
curves along the reaction pathway ($d$ in the inset figure) for the N atom 
approaching the SiO$_2$ network.  The zero distance is the equilibrium 
position of the N atom in the largest interstitial site of $\alpha$-quartz 
where its corresponding total energy is chosen as the zero energy of the 
system.}
\end{figure}

Similarly to NH, we find a spin-dependent diffusion reaction for atomic
N with the SiO$_2$ network. According to our results, the N atom in the ground
quartet state does not react with the network remaining at interstitial sites,
whereas, in the doublet state, the N atom is incorporated between a Si-O bond
forming a structure similar to the peroxyl bridge of 
oxygens\cite{hamann}.
Therefore, we examine the possibility that the N atom, initially in the quartet
state, changes the spin configuration to the doublet one, resulting in its 
incorporation into the oxide.
Following the same procedure described above for NH, we compute the quartet 
and doublet PES for the N atom approaching the SiO$_2$ network. 
Our results are shown in Fig.~\ref{f5}. 
From Eq.~\ref{e2} we obtain the probability for the spin conversion using for 
$V$ the quartet-to-doublet spin-orbit matrix element of the N atom 
($^{4}\!P\,\rightarrow\,^{2}\!P$), obtained from spectroscopy data of 
13.8~cm$^{-1}$. Thus, we find a quartet-to-doublet conversion probability 
$P_{qd}$ = 1.6\,$\times$\,10$^{-5}$.
This probability corresponds to a single passage of the N atom through 
the crossing point. To estimate the number of spin-conversion events of
thermally activated N atoms approaching the SiO$_2$ network, we calculate
the ratio between the rate of N following the quartet diffusion PES
($\Gamma_q$) and the rate of N changing from doublet to singlet PES at the
crossing point ($\Gamma_{qd}$). To calculate $\Gamma_q$ we use Eqs.~\ref{e3} 
substituting $\Delta E_t$ by the energy barrier for the quartet diffusion PES
obtained from Fig.~\ref{f2}(b) ($\Delta E_q$\,=\,0.6~eV). 
$\Gamma_{qd}$ was estimated by  Eqs.~\ref{e4} substituting $P_{ts}$ by 
$P_{qd}$ and using $\Delta E_p$\,=\,0.6~eV, the difference between the 
energies at the interstitial site and at the quartet-doublet crossing 
point, obtained from Fig.~\ref{f5}. For a temperature of 800 $^\circ$C, we 
find that $\Gamma_q/\Gamma_{qd}\simeq6\!\times\!10^{4}$. 
This results indicates that the N atom will most likely proceed along the 
quartet PES, i.e., it diffuses through the SiO$_2$ essentially without 
reacting with the network. 

Although very unlikely, the reaction of a doublet N with the oxide 
due to the spin exchange may also occur. Again, following the same procedure
described for NH, we have three possibilities for the doublet N 
diffusion-reaction mechanism once it is incorporated into the oxide: 
(i) {\it The incorporated N diffuses through the SiO$_2$ network by a 
hopping mechanism}. From Eq.~\ref{e3}, substituting $\Delta E_t$ by the
a calculated hopping energy barrier ($\Delta E_h$\,=\,1.8~eV), we find a 
reaction rate of $\Gamma_h$\,=\,$\Gamma_0$(3.5$\times$10$^{-9}$). 
(ii) {\it N is released from the network following the doublet PES}. From
Eq.~\ref{e5}, substituting $\Delta E_s$ by the the doublet PES energy barrier,
obtained from Fig.~\ref{f5} ($\Delta E_d$\,=\,0.9~eV), and $P_{ts}$ by 
$P_{qd}$, we find $\Gamma_d$\,=\,$\Gamma_0$(5.9$\times$10$^{-5}$). 
(iii) {\it N is released from the network changing from the doublet to the 
quartet PES at the crossing point}. From Eq.~\ref{e4}, substituting $P_{ts}$
by $P_{qd}$ and using $\Delta E_p$\,=\,0.6~eV, the difference between the
energies at the interstitial site and at the quartet-doublet crossing point,
obtained from Fig.~\ref{f5}, we find
$\Gamma_{dq}$\,=\,$\Gamma_0$(2.4$\times$10$^{-8}$). 
Therefore, if a N atom is incorporated into the oxide by a quartet-to-doublet 
spin exchange, a rare event, it will be released from the network following
the singlet PES, according to the reaction (ii). 
However, as the incorporation reaction of the doublet N is exothermic, 
it will be reincorporated in the network if the timescale for relaxation
back to the quartet ground state is long enough.
Finally, we find that the other two possible reactions are very unlikely as 
can be verified by the ratios
$\Gamma_d/\Gamma_h\simeq1.7\!\times\!10^{4}$ and
$\Gamma_d/\Gamma_{dq}\simeq2.5\times\!10^{3}$. 

\section{conclusions}

We conclude that because of the ionic character of SiO$_2$, charged 
species are trapped by the network, where positive (negative) species 
tend to attach to oxygen (silicon) atoms, forming strong bonds with
binding energies as high as 3~eV. This suggests a mechanism for the N 
incorporation in bulk SiO$_2$ observed in plasma assisted method, which 
would be associated to charged species surviving the first stages of the
incorporation reactions occurring at the surface. Of course, these
charged species may also transfer the charge to the network becoming neutral.
According to our results, neutral species do not react with the network 
remaining at interstitial sites, therefore, they would be diffusing species 
in SiO$_2$ able to reach the Si/SiO$_2$ interface. 

We estimate the diffusivities of neutral species through $\alpha$-quartz.
Our results show that NH and atomic N have increasing diffusivities with 
temperature, and the highest one among the nitriding species, suggesting 
that they would be efficient for near interface nitridation. This result
is in good agreement with the finding of high N concentration at the 
Si/SiO$_2$ interface obtained by thermal processing in NH$_3$.
On the other hand, NO and N$_2$ show relatively constant diffusivities
in $\alpha$-quartz for the same range of temperatures.
This result is also in good agreement with the finding of relatively low N
concentration at the Si/SiO$_2$ interface obtained by thermal processing
in N$_2$O, which is the main source of NO and N$_2$ species.

We also study the incorporation reaction of NH and atomic N with the SiO$_2$
network driven by a spin exchange mechanism. We find that, for a typical 
annealing temperature ($T$\,=\,800 $^{\circ}$C), a NH molecule diffusing
through the triplet PES will have, on average, one in forteen possibilities 
to be incorporated into the network by a triplet-to-singlet conversion.
The incorporated singlet NH will most likely return to the triplet PES after 
an inverse spin conversion. 
We also find that atomic N may be incorporated into the network by a
quartet-to-doublet conversion while diffusing through the quartet PES,
however, only an average of one in 6$\times$10$^{4}$ events will be successful.
If this rare event occurs, the incorporated N will be released from the network 
following the doublet PES and, subsequently, reincorporated into the network 
due to the doublet exothermic process. However, this mechanism will be valid
if the timescale for the relaxation to the quartet ground state is long enough.

\acknowledgments
This work was supported by FAPESP and CNPq. We also would like to thank to
CENAPAD-SP for computer time.

\end{document}